\documentclass[journal]{IEEEtran}

\usepackage{cite}
\usepackage[caption=false,font=footnotesize]{subfig}
\usepackage{graphicx}
\usepackage[cmex10]{amsmath}
\usepackage{amsfonts,amssymb,amsthm}
\usepackage{mathrsfs}
\usepackage{enumerate}
\usepackage{color}
\usepackage[makeroom]{cancel}
\usepackage[utf8]{inputenc}
\usepackage[english]{babel}
\usepackage{verbatim}
\usepackage{hyperref}
\theoremstyle{definition}
\newtheorem{dfn}{Definition}

\newtheorem{theorem}{Theorem}
\newtheorem{lemma}{Lemma}
\newtheorem{corollary}{Corollary}
\newtheorem*{remark}{Remark}

\author{
Amjad~Saeed~Khan and Ioannis~Chatzigeorgiou
	\thanks{A. S. Khan and I. Chatzigeorgiou are with the School of Computing and Communications, Lancaster University, Lancaster, United Kingdom (e-mail: \{a.khan9, i.chatzigeorgiou\}@lancaster.ac.uk).}
}
\begin{document}
\font\myfont=cmr12 at 22pt
\title{{\myfont Improved bounds on the decoding failure probability of network coding over multi-source multi-relay networks}}
\maketitle
\begin{abstract}
This letter considers a multi-source multi-relay network, in which relay nodes employ a coding scheme based on random linear network coding on source packets and generate coded packets. If a destination node collects enough coded packets, it can  recover the packets of all source nodes. The links between source-to-relay nodes and relay-to-destination nodes are modeled as packet erasure channels. Improved bounds on the probability of decoding failure are presented, which are markedly close to simulation results and notably better than previous bounds. Examples demonstrate the tightness and usefulness of the new bounds over the old bounds.
\end{abstract} 
\section{Introduction}
\label{intro_section}
The exploitation of cooperative diversity and the inclusion of network coding in multi-source multi-relay networks, in order to achieve excellent performance and high diversity gain, has attracted the interest of the research community. For example, it has been demonstrated in~\cite{Impact_NC} that the use of network coding combined with cooperative diversity not only increase network reliability but also improve network throughput. Moreover, it has been shown in~\cite{CoCast_NC} and~\cite{katti2008network} that they can be significantly useful for wireless networks with disruptive channel and connectivity conditions.

This letter considers linear network coding over a multi-source multi-relay network, where $N$ source nodes are supported by $M$ relay nodes for the delivery of packets over packet erasure channels. To the best of our knowledge, an exact expression for the probability of decoding failure at a destination is not available but an effort has been made in~\cite{Seong_LTR}, in which the author derives upper and lower bounds. However, the bounds presented in~\cite{Seong_LTR} are tight only for a certain range of parameters, including erasure probabilities, the values of $N$, $M$ and the size of the finite field. As shown in Section~\ref{sec:result_sec} of this letter, the existing upper bound is poor for a large number of source nodes and for large finite fields. Moreover, the existing lower bound is independent of the field size and is loose for small finite fields and low erasure probabilities. 

The motivation for this work is to derive improved bounds on the probability of decoding failure. To this end, the main contributions of this letter can be summarized as follows: (i) we have reformulated the problem statement in order to associate the probability of decoding failure with the probability of attaining singular sparse random matrices, and (ii) we have revisited the expressions for upper and lower bounds in~\cite{Seong_LTR} and obtained alternative expressions, which are better than the previous bounds for any field size and any value of $N$ and $M$. Furthermore, the proposed lower bound incorporates the effect of the field size in contrast to the previous lower bound.
\section{System Model}
\label{sec:system}
\begin{figure}[t]
\centering
\includegraphics[width=0.59\columnwidth]{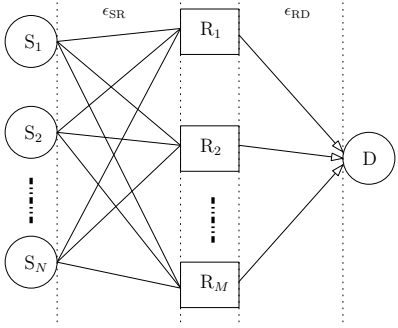}
\caption{A network consisting of $N$ source nodes, $M\geq N$ relay nodes and a destination $\mathrm{D}$. The packet erasure probability of a source-to-relay link and a relay-to-destination link is represented by $\epsilon_{\mathrm{SR}}$ and $\epsilon_{\mathrm{RD}}$, respectively.} 
\vspace{-4mm}
\label{fig:system model}
\end{figure}

We consider a system with $N$ source nodes and $M$ relay nodes, $\{\mathrm{S}_1,\mathrm{S}_2,\hdots,\mathrm{S}_N\}$ and $\{\mathrm{R}_1,\mathrm{R}_2,\hdots,\mathrm{R}_M\}$, respectively, as shown in Fig.~\ref{fig:system model}, where $M\geq N$. Each source node $\mathrm{S}_i$ has a packet $x_i$ to transmit to a destination $\mathrm{D}$ via $M$ relay nodes. No source-to-destination links are assumed. The links connecting source-to-relay and relay-to-destination nodes are modeled as independent packet erasure channels characterized by erasure probability $\epsilon_{\mathrm{SR}}$ and $\epsilon_{\mathrm{RD}}$, respectively. 

The communication process is split into two phases. In the first phase, all the source nodes transmit their information packets simultaneously to the relay nodes over orthogonal broadcast channels. In the second phase, each relay node generates a coded packet by randomly combining the successfully received packets from the $N$ source nodes. The $M$ coded packets are then forwarded to the destination $\mathrm{D}$ over orthogonal channels. The coded packet $y_i$, which is transmitted by the $i^\mathrm{th}$ relay node, can be expressed as $y_i=\sum_{j=1}^{N}c_{i,j}x_j$, where $c_{i,j}$ is a coding coefficient selected independently at random over a finite field $F_q$ of size $q$. 
Because of the link condition $\epsilon_{\mathrm{SR}}$ between the source node $\mathrm{S}_j$ and the relay node $\mathrm{R}_i$, each relay node receives packets from different source nodes. In contrast to~\cite{Multicast_capacity_2} where coding coefficients are chosen uniformly at random, our system model imposes that the zero coefficient is assigned to erased packets and the remaining $q-1$ non-zero coefficients are selected uniformly at random by each relay for successfully received packets. Consequently, the coding coefficient distribution is given by
\begin{equation}
\begin{split}
\label{eq:uniform_distribution}
P[c_{i,j}=t]&=\left\{
	\begin{array}{ll}
		\!\epsilon_{\mathrm{SR}},&\!\!\!\mbox{if }t = 0\\[0.5em]
		\displaystyle\!\!\frac{1-\epsilon_{\mathrm{SR}}}{q-1}, &\!\!\!\mbox{if }t \in F_q\setminus\{0\}
	\end{array}
\right.
\end{split}
\end{equation}
where $0\leq \epsilon_{\mathrm{SR}}\leq1$. For a given relay node $i$, the sequence $c_{i,1},\hdots, c_{i,N}$ forms a row vector, which is known as the coding vector of the coded packet $y_i$. As is commonly assumed in network coding~\cite{Fitzek_ICC}, coding vectors are transmitted along with the corresponding coded packets. When the destination $\mathrm{D}$ receives $N$ linearly independent coded packets, the packets of all source nodes can be recovered. Transmission of source packets over erasure channels and random linear coding at relay nodes is analogous to sparse random linear Network Coding (NC), which uses sparse random matrices~\cite{Blomer_1997,C-Cooper}. Based on the work of Bl$\ddot{\text{o}}$mer~\cite{Blomer_1997} and Cooper~\cite{C-Cooper}, this paper derives improved upper and lower bounds on the probability that the destination will fail to recover the source packets. 
\section{Preliminary Results and Former Bounds on the Probability of Decoding Failure}
Consider a matrix $\mathbf{A} \in F_q^{M \times N}$, whose elements are the coding coefficients $c_{i,j}$ such that the $i^\mathrm{th}$ row of $\mathbf{A}$ represents the coding vector associated with the $i^\mathrm{th}$ coded packet received by the destination $\mathrm{D}$. The destination can recover the packets of the $N$ source nodes if and only if $rank(\mathbf{A})=N$. Thus, the decoding failure probability at the destination $\mathrm{D}$ can be defined as $\resizebox{0.21\textwidth}{!}{$P_{\mathrm{fail}}\!:=\!\mathrm{Pr}\{rank(\mathbf{A})\!<\!N\}$}$. It is related to the linear dependence of the vectors of matrix $\mathbf{A}$ and is defined as:
\begin{dfn} The vectors of matrix $\mathbf{A}\in F_q^{M \times N}$ are said to be \emph{linearly dependent} if and only if there exists a column vector \mbox{$\mathbf{x}\in F^{N\times 1}_q\backslash \{\mathbf{0}\}$} such that
\begin{equation}
\mathbf{A}\mathbf{x}=\mathbf{0}.
\end{equation}
\end{dfn}
When there is no packet loss between the relay-to-destination channels, i.e., $\epsilon_{\mathrm{RD}}=0$, the probability that the elements of the $i^\mathrm{th}$ row of matrix $\mathbf{A}$ add up to zero, i.e., $c_{i,1}+c_{i,2}+\hdots+c_{i,N}=0$, is given by~\cite{Blomer_1997}
\begin{equation}
\gamma_N=q^{-1}+(1-q^{-1})(1-\frac{1-\epsilon_{\mathrm{SR}}}{1-q^{-1}})^N.
\end{equation}
Taking into account that matrix $\mathbf{A}$ consists of $M$ rows, the probability $\mathrm{Pr}(\mathbf{A}\mathbf{x}=\mathbf{0})$ can be obtained as
\begin{equation}
\resizebox{0.4376\textwidth}{!}{$
\displaystyle \mathrm{Pr}(\mathbf{A}\mathbf{x}=\mathbf{0})=\gamma_N^{M}=\big(q^{-1}+(1-q^{-1})(1-\frac{1-\epsilon_{\mathrm{SR}}}{1-q^{-1}})^N\big)^M. 
$}
\end{equation}
The expected number of decoding failures at the destination $\mathrm{D}$ is given by the following theorem, which is a straightforward adaptation of~\cite[Theorem~3.3]{Blomer_1997},~\cite[Theorem~3]{C-Cooper} to the system model under consideration.   
\begin{theorem}
\label{pro:theorem_1}
For a linear network coding scheme over $N$ source nodes, $M\geq N$ relay nodes and a single or multiple destinations, which are interconnected by links characterized by packet erasure probabilities $0\leq\epsilon_{\mathrm{SR}}\leq 1$ and $\epsilon_{\mathrm{RD}}=0$, the expectation of the decoding failures can be obtained as 
\begin{equation}
\label{eq:theorem_eq_1}
 \mu_0(N,M)\!=\!{E}(\mathbf{A}\mathbf{x}=\mathbf{0})\!=\!\frac{1}{q-1}\sum_{w=1}^{N}\binom{N}{w}(q-1)^w\gamma_w^{M}
\end{equation}
where $\mathbf{A}\in F_q^{M\times N}$ is the coding matrix at a destination.
\end{theorem}
Following the same line of reasoning, a direct extension of~\eqref{eq:theorem_eq_1} for $\epsilon_{\mathrm{RD}}\geq 0$ has been made in~\cite[Theorem 1]{Seong_LTR} and was used to upper bound the probability of decoding failure.
\begin{corollary}
The probability of decoding failure at a destination is bounded from above as:
\begin{equation}
\label{eq:seong_eq_1}
\displaystyle P_\mathrm{fail}\leq\!\frac{1}{q-1}\!\!\sum_{w=1}^{N}\!\!\binom{N}{w}\!(q-1)^w\!\big [\epsilon_{\mathrm{RD}}+(1-\epsilon_{\mathrm{RD}})\gamma_w\big ]^M 
\end{equation}
where $N$ is the number of source nodes, $M\geq N$ is the number of relay nodes and $\epsilon_{\mathrm{SR}}$, $\epsilon_{\mathrm{RD}}$ represent the packet erasure probabilities between the network nodes.
\end{corollary}
However,~\eqref{eq:seong_eq_1} is only tight for limited values of erasures $\epsilon_{\mathrm{SR}}$ and $\epsilon_{\mathrm{RD}}$, depending on $N$, $M$ and $q$. In particular, the upper bound takes values greater than 1 when either the field size is big or the difference between the number of source and relay nodes is small. This disparity between the probability of decoding failure and the upper bound will be demonstrated in Section~\ref{sec:result_sec}. In an effort to improve the tightness of~\eqref{eq:seong_eq_1}, Seong \emph{et al.} proposed the selection of the minimum value between the upper bound in~\eqref{eq:seong_eq_1} and 1~\cite{Seong_TCOM}.
A lower bound on the probability of decoding failure has also been obtained by Seong in~\cite[Theorem~2]{Seong_LTR}: 
\begin{theorem}
\label{pro:theorem_2}
Consider a network comprising $N$ source nodes and $M\geq N$ relay nodes, assume that  links are modeled as packet erasure channels with erasure probabilities $\epsilon_{\mathrm{SR}}$ and $\epsilon_{\mathrm{RD}}$, and let $\mathbf{A}\in F_q^{M\times N}$ be the coding matrix at a destination node. The probability of decoding failure $P_\mathrm{fail}$ is lower bounded by 
\begin{equation}
\label{eq:theorem_eq_2}
\begin{split}
P_\mathrm{fail}\geq&\sum_{k=1}^{N}\binom{N}{k}\big ((\epsilon_{\mathrm{SR}}+\epsilon_{\mathrm{RD}}-\epsilon_{\mathrm{SR}}\epsilon_{\mathrm{RD}})^{M}\big )^{k}\\
&\times(1-(\epsilon_{\mathrm{SR}}+\epsilon_{\mathrm{RD}}-\epsilon_{\mathrm{SR}}\epsilon_{\mathrm{RD}})^{M})^{N-k}.
\end{split}
\end{equation}
\end{theorem}
The bounds in~\eqref{eq:seong_eq_1} and~\eqref{eq:theorem_eq_2} are used in~\cite{Seong_TCOM} and~\cite{disaster_conf}. For example,~\eqref{eq:seong_eq_1} is employed in~\cite{disaster_conf} to evaluate the performance gains introduced by linear NC  in a practical network architecture for emergency communications. However, the following section will derive new bounds, which are considerably tighter than the previous bounds and can significantly improve the quality and accuracy of results presented in the literature. 
\section{Improved Bounds on the Probability of Decoding Failure}
\subsection{Upper Bound}
For $\epsilon_{\mathrm{RD}}=0$, an upper bound on the decoding failure probability can be obtained by extending and adapting~\cite[Theorem 6.3]{Blomer_1997} as follows:
\begin{lemma}
\label{pro:lemma_1}
Let $\mathbf{A}\in F_q^{M\times N}$ be the coding matrix at a destination node of a network consisting of $N$ source nodes and $M$ relay nodes. If the internode erasure probabilities are $0\leq \epsilon_{\mathrm{SR}}\leq 1$ and $\epsilon_{\mathrm{RD}}=0$, the probability of decoding failure is upper bounded by 
\begin{equation}
\label{eq:lemma_1_eq}
\eta_\mathrm{max}(N,M)=1-\prod_{i=1}^{N}(1-\beta_{\max}^{M-i+1})
\end{equation}
where $\displaystyle\beta_{\max}=\max(\epsilon_{\mathrm{SR}},\frac{1-\epsilon_{\mathrm{SR}}}{q-1})$.\\
\end{lemma}
\begin{IEEEproof}
Let us assume that the first $i-1$ columns of $\mathbf{A}$, denoted by $\mathbf{A}_1,\mathbf{A}_2,\hdots,\mathbf{A}_{i-1}$, are linearly independent. This implies that by using elementary column operations, matrix $\mathbf{A}$ can be transformed into a matrix that contains an
\mbox{$(i-1)\times(i-1)$} identity matrix. Without loss of generality, let us assume that the first $i-1$ rows form the identity matrix. The columns of this matrix represent the basis for the vector space spanned by $\mathbf{A}_1,\mathbf{A}_2,\hdots,\mathbf{A}_{i-1}$. Therefore, the probability that $\mathbf{A}_i$ is linearly independent from $\mathbf{A}_1,\mathbf{A}_2,\hdots,\mathbf{A}_{i-1}$ depends only on the last $M\!-\!i\!+\!1$ elements of $\mathbf{A}_i$. This probability is lower bounded by $1-\beta_{\max}^{M-i+1}$, where $\beta_{\max}$ specifies the maximum probability of obtaining an element from $F_q$.
Hence, matrix $\mathbf{A}$ contains an $N\times N$ non-singular matrix with probability at least $\resizebox{0.15\textwidth}{!}{$\prod_{i=1}^{N}(1-\beta_{\max}^{M-i+1})$}$. As a result, the probability that matrix $\mathbf{A}$ does not contain an invertible matrix and, consequently, a decoding failure will occur is upper bounded by subtracting this product from one, which completes the proof.
\end{IEEEproof}
Lemma~\ref{pro:lemma_1} will be used to obtain a tighter upper bound on $P_{\mathrm{fail}}$. Before we invoke it, we shall first revisit~\eqref{eq:seong_eq_1} and rewrite it as:
\begin{equation}
\label{eq:lemma_2_eq}
P_\mathrm{fail}\leq\sum_{r=0}^{M}\binom{M}{r}\epsilon_{\mathrm{RD}}^{M-r}(1-\epsilon_{\mathrm{RD}})^r{\mu_0}(N,r).
\end{equation}
This change is possible if $\ [\epsilon_{\mathrm{RD}}+(1-\epsilon_{\mathrm{RD}})\gamma_w\ ]^M$ is expanded into a sum, as per the binomial theorem.
\begin{theorem}
\label{pro:theorem_3}
For a network coding scheme over multi-source multi-relay networks, composed of $N$ source nodes and $M$ relay nodes with packet erasures $\epsilon_{\mathrm{SR}}$ and $\epsilon_{\mathrm{RD}}$, the probability of decoding failure is upper bounded by 
\begin{equation}
\label{eq:theorem_3_eq}
\resizebox{0.4366\textwidth}{!}
{$
\displaystyle
P_\mathrm{fail}\!\leq\!\sum_{r=0}^{M}\!\binom{M}{r}\epsilon_{\mathrm{RD}}^{M-r}(1-\epsilon_{\mathrm{RD}})^r\!\min\{\eta_\mathrm{max}(N,r),\mu_0(N,r)\}.
$}
\end{equation}
\end{theorem}
\begin{IEEEproof}
As inferred from~\eqref{eq:lemma_2_eq}, the number of packet deliveries by the relays follows the binomial distribution. If we employ Theorem~\ref{pro:theorem_1} and Lemma~\ref{pro:lemma_1} on the number of received coded packets $r$, a tight upper bound can be obtained by taking the minimum of outcomes and multiply with the probability distribution of $r$. Summing the resultant quantity gives~\eqref{eq:theorem_3_eq}, which concludes the proof. 
\end{IEEEproof}
\begin{remark} It is worth noting that the upper bound is not simply the minimum between two cumulative probability distributions (CDFs), that is, the right-hand of~\eqref{eq:seong_eq_1} and the CDF of~\eqref{eq:lemma_1_eq} for \textit{all} possible numbers of relay nodes. Instead, the right hand of~\eqref{eq:seong_eq_1} has been rewritten in the form of~\eqref{eq:lemma_2_eq}, which enabled us to identify the minimum between $\mu_0$ and $\eta_\mathrm{max}$ for \textit{each} possible number of relay nodes, and use it in the computation of the CDF shown in~\eqref{eq:lemma_2_eq}.
\end{remark}
\subsection{Lower Bound}
The bound that was derived in~\cite[Theorem 6.3]{Blomer_1997} was extended to an upper bound on the probability that an $M\times N$ matrix $\mathbf{A}$ does not contain an invertible $N\times N$ matrix in Lemma~\ref{pro:lemma_1}. The same approach can be followed to obtain a lower bound as follows:
\begin{lemma}
\label{pro:lemmaupdate_2}
Let $\mathbf{A}\in F_q^{M\times N}$ be the coding matrix at a destination of a network consisting of $N$ source nodes and $M$ relay nodes. If the internode erasure probabilities are $0\leq \epsilon_{\mathrm{SR}}\leq 1$ and $\epsilon_{\mathrm{RD}}=0$, the probability of decoding failure is lower bounded by 
\begin{equation}
\label{eq:lower_bound_eq}
\eta_\mathrm{min}(N,M)=1-\prod_{i=1}^{N}(1-\beta_{\min}^{M-i+1})
\end{equation}
where $\displaystyle\beta_{\min}=\min(\epsilon_{\mathrm{SR}},\frac{1-\epsilon_{\mathrm{SR}}}{q-1})$.
\end{lemma}
\begin{IEEEproof}
The proof follows exactly the same line of reasoning as that of Lemma~\ref{pro:lemma_1}.
\end{IEEEproof}
An improved lower bound on $P_{\mathrm{fail}}$ can be obtained if the right-hand side of (8) is denoted by $P_{0}(N,M)$ for $\epsilon_{\mathrm{RD}}=0$, that is  
\begin{equation}
\label{eq:lemma_3_eq}
P_{0}(N,M)=\sum_{k=1}^{N}\binom{N}{k}(\epsilon_{\mathrm{SR}}^{M})^k(1-\epsilon_{\mathrm{SR}}^{M})^{N-k}\\
\end{equation}
and then combined with~\eqref{eq:lower_bound_eq} in Lemma~\ref{pro:lemmaupdate_2}. In particular:
\begin{theorem}
\label{pro:theorem_4}
For a linear network coding scheme over $N$ source nodes and $M\geq N$ relay nodes, let $\epsilon_{\mathrm{SR}}$ and $\epsilon_{\mathrm{RD}}$ be the packet erasure probabilities of the internode links. The probability of decoding failure is lower bounded by
\begin{equation}
\label{eq:theorem_4_eq}
\resizebox{0.4366\textwidth}{!}
{$
\displaystyle
P_\mathrm{fail}\!\geq\!\sum_{r=0}^{M}\binom{M}{r}\epsilon_{\mathrm{RD}}^{M-r}(1-\epsilon_{\mathrm{RD}})^r\!\max\{\eta_\mathrm{min}(N,r),P_{0}(N,r)\}.
$}
\end{equation}
\end{theorem}
\begin{IEEEproof}
In contrast to Theorem~\ref{pro:theorem_3}, here we employ Lemma~\ref{pro:lemmaupdate_2} and~\eqref{eq:lemma_3_eq} on the number of received coded packets $r$, and we select the maximum of outcomes. The rest of the proof follows the same reasoning as that presented in the proof of Theorem~\ref{pro:theorem_3}. 
\end{IEEEproof}
\vspace{-0.30mm}
\section{Results}
\label{sec:result_sec}
This section compares the analytical expressions of the proposed bounds to simulation results. In addition, the proposed upper bound and lower bound, which shall be referred to as \mbox{UB-new} and \mbox{LB-new}, are contrasted with the old bounds represented by~\eqref{eq:seong_eq_1} and~\eqref{eq:theorem_eq_2}, which shall be referred to as \mbox{UB-old} and \mbox{LB-old}. To obtain simulation results, each scenario was run over $10^4$ realizations, failures by the destination to recover the packets of all source nodes were counted, and the decoding failure probability was measured.

\begin{figure}[h]
\includegraphics[width=.975\columnwidth]{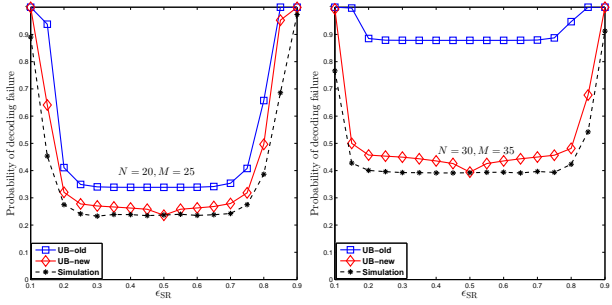}
\vspace{-4.0mm}
\caption{Comparison between simulation results and the theoretical upper bounds obtained from~\eqref{eq:seong_eq_1} and~\eqref{eq:theorem_3_eq} for different values of $N$ and $M$, when $q=2$, $\epsilon_{\mathrm{RD}}=0.1$ and $\epsilon_{\mathrm{SR}} \in [0.1, 0.9]$.} 
\vspace{-2.53mm}
\label{fig:figure_2}
\end{figure}

Fig.~\ref{fig:figure_2} shows numerical results of the upper bounds obtained from~\eqref{eq:seong_eq_1} and~\eqref{eq:theorem_3_eq} and labeled UB-old and UB-new, respectively. We observe that, in contrast to \mbox{UB-old}, \mbox{UB-new} is significantly tighter to the simulated performance. When the number of source nodes and the number of relay nodes increase to $N=30$ and $M=35$, respectively, it can be clearly seen that the \mbox{UB-old} curve moves far away from the simulated curve but the proposed UB-new expression still provides a tight bound. This reveals the fact that \mbox{UB-old} produces a worse approximation error for large values of $N$. 

\begin{figure}[h]
\includegraphics[width=.96\columnwidth]{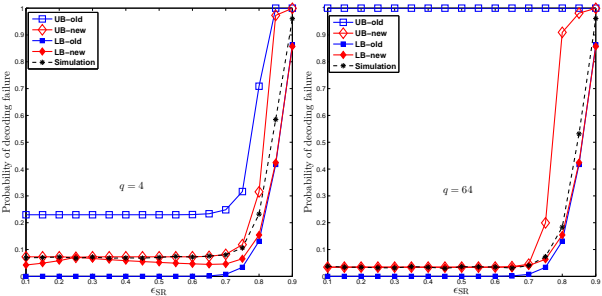}
\vspace{-4.0mm}
\caption{Effect of field size $q$ on network performance and comparison between the proposed bounds and the old bounds for $\epsilon_{\mathrm{SR}}\in[0.1, 0.9]$, when $N=20$, $M=25$ and $\epsilon_{\mathrm{RD}}=0.1$.} 
\vspace{-1.4mm}
\label{fig:figure_3}
\end{figure}
\begin{figure}[t]
\centering
\includegraphics[width=0.959\columnwidth]{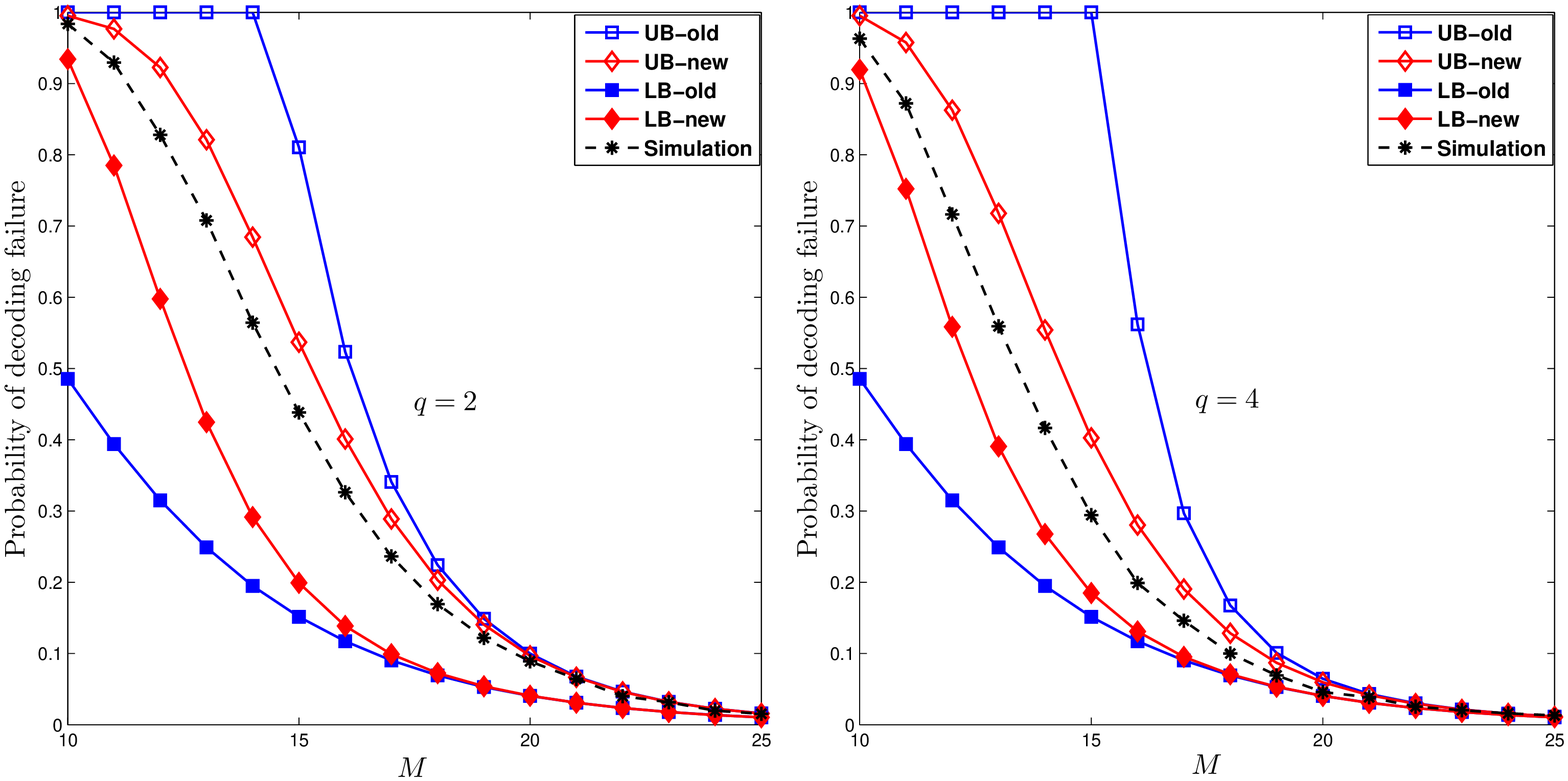}
\vspace{-4.0mm}
\caption{Performance of the network for an increasing number of relays $M$. The proposed bounds and the old bounds have been plotted for $N=10$, $\epsilon_{\mathrm{SR}}=0.7$, $\epsilon_{\mathrm{RD}}=0.2$ and different values of field size $q$.}
\vspace{-2.53mm}
\label{fig:figure_4}
\end{figure}
\begin{figure}[t]
\centering
\includegraphics[width=.972\columnwidth]{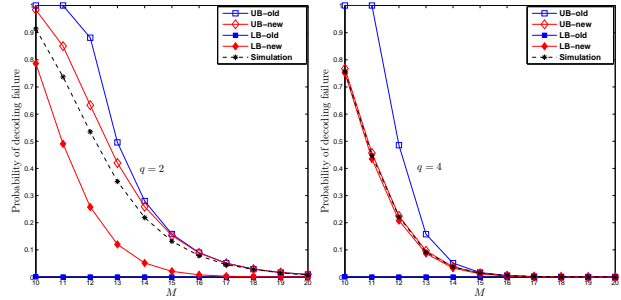}
\vspace{-4.0mm}
\caption{Network performance and comparison between the proposed bounds and the old bounds for $N=10$, an increasing number of relays $M$, \mbox{$\epsilon_{\mathrm{SR}}=0.3$}, $\epsilon_{\mathrm{RD}}=0.1$ and different field size $q$.} 
\label{fig:figure_5}
\end{figure}
Fig.~\ref{fig:figure_3} evaluates the probability of decoding failure for $q=\{4,64\}$, and contrasts the proposed bounds (\mbox{UB-new} and \mbox{LB-new}) with the old bounds (\mbox{UB-old} and \mbox{LB-old}). The figure demonstrates that for $\epsilon_{\mathrm{SR}}\in [0.1,0.7]$, the network experiences only a small probability of decoding failure. Furthermore, the figure shows that \mbox{UB-new} and \mbox{LB-new} are very close to the simulated performance and outperform \mbox{UB-old} and \mbox{LB-old}, respectively. In particular, when $q=64$, \mbox{UB-old} and \mbox{LB-old} are markedly loose while \mbox{UB-new} and \mbox{LB-new} are very tight to the actual simulation results. The performance of the network deteriorates for values of $\epsilon_{\mathrm{SR}}$ greater than 0.75. Moreover it is interesting to notice that, for large values of $q$, the upper bounds deviate from the simulation results and the simulations can be better approximated by the lower bounds. 

Figs.~\ref{fig:figure_4} and~\ref{fig:figure_5} plot the probability of decoding failure versus the number of relays $M$ with $N\!\!=\!\!10$ and $q\!\!=\!\!\{2,4\}$. 
It is evident that the probability of decoding failure decreases with an increasing number of relays and field size. The figures also demonstrate that, when $M\!<\!2N$, \mbox{UB-new} and \mbox{LB-new} are close to the simulated outcomes, compared to \mbox{UB-old} and \mbox{LB-old}, respectively. It follows from~\eqref{eq:theorem_eq_2} that \mbox{LB-old} depends only on the erasures $\epsilon_{\mathrm{SR}}$ and $\epsilon_{\mathrm{RD}}$, and does not depend on the field size $q$, thus shows no improvement for \mbox{$q=4$}. However, LB-new approaches the simulation results, when $q$ increases to 4. For example in Fig.~\ref{fig:figure_5}, when $q=4$ and $M\leq 14$, both \mbox{UB-new} and \mbox{LB-new} are very tight, while \mbox{UB-old} and \mbox{LB-old} are noticeably far from the simulated performance. 
\section{Conclusions}
We presented improved upper and lower bounds on the probability of decoding failure in a multi-source multi-relay network, which employs linear network coding. The proposed analysis for counting failures provided significantly tighter bounds, which outperform existing bounds, derived in~\cite{Seong_LTR}. Several examples, which considered various numbers of source nodes and relay nodes, different field sizes and a range of erasure probabilities, established the shortcomings of the existing bounds and demonstrated the tightness of the proposed improved bounds. Finally, we assert that the proposed bounds can also be used to better estimate the performance of systems employing sparse random linear network coding schemes, presented in the literature.\\
\vspace{-3.90mm}
\bibliographystyle{IEEEtran}
\bibliography{IEEEabrv,IEEE_CommLett_Ref}

\begin{thebibliography}{10}
\providecommand{\url}[1]{#1}
\csname url@samestyle\endcsname
\providecommand{\newblock}{\relax}
\providecommand{\bibinfo}[2]{#2}
\providecommand{\BIBentrySTDinterwordspacing}{\spaceskip=0pt\relax}
\providecommand{\BIBentryALTinterwordstretchfactor}{4}
\providecommand{\BIBentryALTinterwordspacing}{\spaceskip=\fontdimen2\font plus
\BIBentryALTinterwordstretchfactor\fontdimen3\font minus
  \fontdimen4\font\relax}
\providecommand{\BIBforeignlanguage}[2]{{%
\expandafter\ifx\csname l@#1\endcsname\relax
\typeout{** WARNING: IEEEtran.bst: No hyphenation pattern has been}%
\typeout{** loaded for the language `#1'. Using the pattern for}%
\typeout{** the default language instead.}%
\else
\language=\csname l@#1\endcsname
\fi
#2}}
\providecommand{\BIBdecl}{\relax}
\BIBdecl

\bibitem{Impact_NC}
Z.~Ding, Z.~Ma, and K.~K. Leung, ``Impact of network coding on system delay for
  multisource-multidestination scenarios,'' \emph{{IEEE} Trans. Veh. Technol.},
  vol.~59, no.~2, pp. 831--841, Feb. 2010.

\bibitem{CoCast_NC}
J.-S. Park, M.~Gerla, D.~S. Lun, Y.~Yi, and M.~M{\'e}dard, ``Codecast: {A}
  network-coding-based ad hoc multicast protocol,'' \emph{{IEEE} Trans.
  Wireless Commun.}, vol.~13, no.~5, pp. 76--81, Oct. 2006.

\bibitem{katti2008network}
S.~R. Katti, ``Network coded wireless architecture,'' \emph{Ph.D Dissertation,
  Massachusetts Institute of Technology, USA}, Sep. 2008.

\bibitem{Seong_LTR}
J.-T. Seong, ``Bounds on decoding failure probability in linear network coding
  schemes with erasure channels,'' \emph{{IEEE} Commun. Lett.}, vol.~18, no.~4,
  pp. 648--651, Apr. 2014.

\bibitem{Multicast_capacity_2}
T.~Ho, M.~M\'{e}dard, R.~Koetter, D.~R. Karger, M.~Effros, J.~Shi, and
  B.~Leong, ``A random linear network coding approach to multicast,''
  \emph{{IEEE} Trans. Inf. Theory}, vol.~52, no.~10, pp. 4413--4430, Oct. 2006.

\bibitem{Fitzek_ICC}
J.~Heide, M.~V. Pedersen, F.~H. Fitzek, and M.~M{\'e}dard, ``On code parameters
  and coding vector representation for practical {RLNC},'' in \emph{Proc. IEEE
  ICC}, Kyoto, Japan, Jun. 2011.

\bibitem{Blomer_1997}
J.~Bl{\"o}mer, R.~Karp, and E.~Welzl, ``The rank of sparse random matrices over
  finite fields,'' \emph{\!Rand.\,\,\!Struct.\,\,\!Alg.}, vol. \!10, no. \!4,
  pp. \!407--420, Jul. \!1997.

\bibitem{C-Cooper}
C.~Cooper, ``On the distribution of rank of a random matrix over a finite
  field,'' \emph{Rand. Struct. Alg.}, vol.~17, no. 3-4, pp. 197--212, Oct.
  2000.

\bibitem{Seong_TCOM}
J.-T. Seong and H.-N. Lee, ``Predicting the performance of cooperative wireless
  networking schemes with random network coding,'' \emph{{IEEE} Trans.
  Commun.}, vol.~62, no.~8, pp. 2951--2964, Aug. 2014.

\bibitem{disaster_conf}
T.~Do-Duy and M.~A. V{\'a}zquez-Castro, ``Efficient communication over cellular
  networks with network coding in emergency scenarios,'' in \emph{Proc.
  ICT-DM}, Rennes, France, Nov. 2015.

\end{thebibliography}
\end{document}